\documentclass[preprint,12pt,authoryear,nopreprintline]{elsarticle}

\usepackage[T1]{fontenc}
\usepackage{lmodern}
\usepackage{amssymb}
\usepackage{amsmath}
\usepackage{graphicx}
\usepackage{booktabs}
\usepackage{tabularx}
\usepackage{array}
\usepackage{caption}
\usepackage{xcolor}
\usepackage{microtype}
\usepackage[hyphens]{url}
\usepackage[hidelinks,breaklinks]{hyperref}
\newcolumntype{Y}{>{\raggedright\arraybackslash}X}
\newcolumntype{P}[1]{>{\raggedright\arraybackslash}p{#1}}

\graphicspath{{figures/}}

\journal{Journal of Systems and Software}

\begin{document}
\begin{frontmatter}

\title{Decomposing Browser Pipeline Architectures for DOM-Sourced Particle Effects:
Worker Offload, WebGL, and WebAssembly}

\author[aut]{Hossein Asadi\corref{cor1}}
\ead{hossein.asadi@aut.ac.ir}
\cortext[cor1]{Corresponding author}
\affiliation[aut]{
  organization={Amirkabir University of Technology},
  addressline={Department of Computer Engineering},
  city={Tehran},
  country={Iran}}

\begin{abstract}
Local optimization does not necessarily yield end-to-end optimization in layered
browser architectures.
Teams often treat Web Workers, WebGL, and WebAssembly as interchangeable ways to
``make it faster,'' yet each lever targets a different layer.
We present a controlled architectural decomposition---using DOM-sourced particle
pipelines as a concrete workload---of five \emph{particle} pipelines (P1--P5), with an
additional non-particle CSS-layer baseline (P0), that isolates thread placement,
renderer choice, and simulation backend.
Using a reproducible harness we measure interactive pacing, high-load end-to-end
stress, and a simulation-only microbenchmark, plus same-host cross-browser /
dual-GPU-class replication (Firefox~153+Intel UHD; Chrome~138+NVIDIA NVK) and an
independent second-host slice on Google Colab (Chrome~150, Tesla~T4, $n{=}5$).
Four results stand out.
(1)~Worker offload improves interactive pacing on paper-primary Chrome
($\approx$144 vs $\approx$52\,FPS; Cliff's $\delta{=}1$, $n{=}10$).
(2)~AssemblyScript speeds the Chrome update kernel by about $1.5$--$1.6\times$
on the primary host and $\approx$1.85$\times$ on Colab T4 (sim-only).
(3)~Under $\sim$250k WebGL particles the Chrome sim-only WASM win need not raise
product FPS (P5$\le$P4 on primary/Firefox; P5$\approx$P4 on Colab T4); in-worker
CPU phase timers show simulate still dominates the accounted budget, so the
non-translation is not a simple ``draw dominates CPU'' story.
(4)~Renderer ranking and absolute margins vary across browser/GPU/host
configurations.
The contribution is bottleneck-aware architectural measurement: identify the dominant
layer and test whether a layer win propagates to user-visible FPS.
\end{abstract}


\end{frontmatter}

\section{Introduction}
\label{sec:intro}

Browser applications increasingly assemble performance-critical paths from interchangeable
platform levers---Web Workers, GPU canvas APIs, and WebAssembly---and treat each lever as
a general ``make it faster'' upgrade.
For software engineering this is an architectural claim, not a style tip: each lever
optimizes a different layer, and a layer-local win need not improve end-to-end,
user-visible performance.
Empirical software-engineering work has long warned that assumed best practices and
engine-dependent optimizations must be measured, not taken on faith; browser stacks now
raise the same validation problem at the architecture level.

\begin{figure}[t]
\centering
\includegraphics[width=0.95\linewidth]{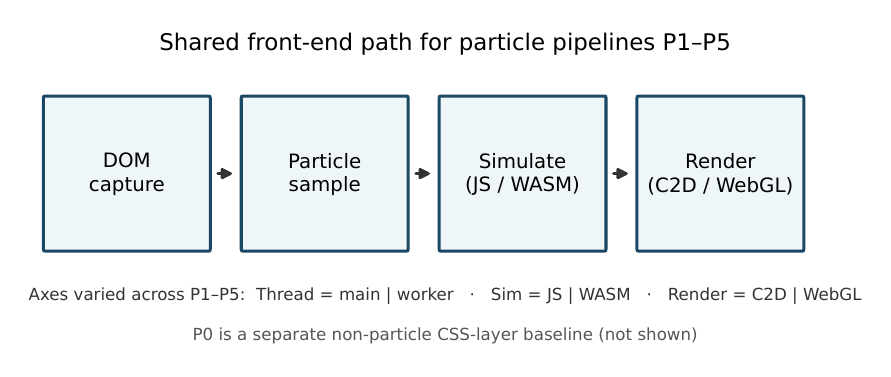}
\caption{High-level path shared by the \emph{particle} pipelines P1--P5.
Architectural axes (thread, simulation backend, renderer) are varied across P1--P5;
P0 is a separate non-particle CSS-layer baseline.}
\label{fig:architecture}
\end{figure}

We study that problem in a concrete, controllable case: DOM-sourced particle dissolve
pipelines (capture a live DOM subtree, sample particles, simulate, render;
Fig.~\ref{fig:architecture}).
The dissolve domain is a \emph{workload}, not the scientific contribution.
It matters because it forces the full stack---DOM capture, CPU simulation, and
Canvas2D/WebGL drawing---into one measurable product path that libraries actually ship.

\begin{figure}[t]
\centering
\includegraphics[width=0.98\linewidth]{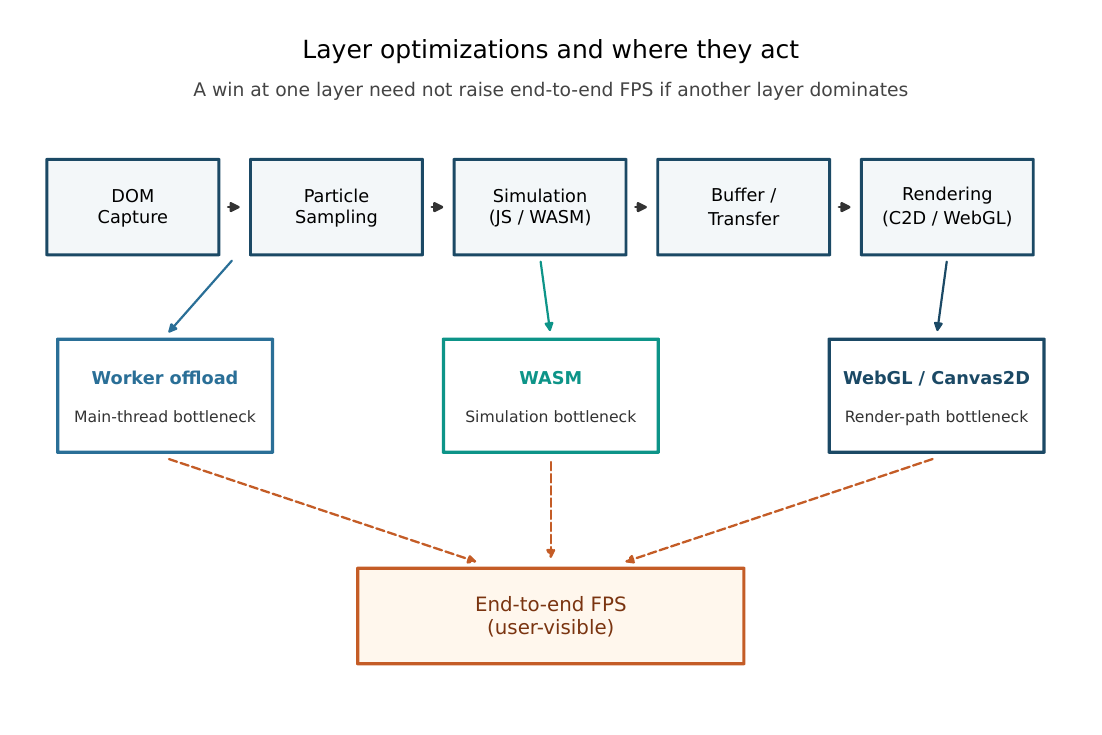}
\caption{Where each lever acts.
Worker targets main-thread contention; WASM targets the simulation kernel;
Canvas2D/WebGL target the render path.}
\label{fig:bottleneck}
\end{figure}

This paper asks a systems-measurement question: \emph{when does a layer optimization
fail to become a product win?}
We isolate thread placement, renderer choice, and simulation backend across five
particle pipelines (P1--P5), with P0 retained only as a non-particle CSS-layer baseline.
P5 is a full Worker + AssemblyScript + WebGL2 stack that mirrors a production-style
effect library, so the comparison stays grounded in shipped architecture rather than a
toy microdemo.

\paragraph{Research questions}
\begin{itemize}
\item \textbf{RQ1.} Does a Worker+WASM+WebGL pipeline improve effect FPS and
frame-time p95 relative to a main-thread Canvas2D particle baseline?
\item \textbf{RQ2.} How do those gaps change as particle density or count increases?
\item \textbf{RQ3.} How does DOM-capture latency grow with fixture complexity?
\item \textbf{RQ4.} Under which measurement regimes do Worker, WebGL, and WASM
contribute to---or fail to translate into---end-to-end performance gains?
\end{itemize}

\paragraph{Contributions}
\begin{enumerate}
\item A \textbf{controlled architectural decomposition} (thread $\times$ simulation
$\times$ renderer) that attributes gains and non-gains to individual browser levers
instead of treating Worker, WebGL, and WASM as interchangeable stack upgrades.
\item A \textbf{three-regime measurement method} (interactive pacing, end-to-end
stress, simulation-only) that separates layer-local evidence from product-FPS claims,
plus same-host cross-browser / dual-GPU-class replication and one independent Colab
Tesla~T4 second-host slice.
\item \textbf{Empirical evidence}---with effect sizes and claim boundaries---that
interactive wins are dominated by worker offload on paper-primary Chrome, that
AssemblyScript speeds Chrome-class update kernels by about $1.5$--$1.85\times$, and that
under WebGL-heavy load those kernel gains need not appear as a clear end-to-end FPS win.
\end{enumerate}

\paragraph{Non-goals}
We do not contribute a new dissolve aesthetic, a new browser API, or a universal ranking
of Worker/WebGL/WASM.
The intended generalization is methodological: bottleneck-aware evaluation of layered
browser architectures.

\section{Related Work}
\label{sec:rw}

We organize related work by theme.
Standards docs and engineering blogs define APIs; \emph{research-gap} claims below cite
peer-reviewed empirical studies.

\subsection{Browser performance and empirical web engineering}
Page-load profilers such as WProf show that computation---not only the
network---often lies on the critical path~\citep{wang2013wprof}, and mobile browsers
can be even more compute-bound~\citep{nejati2016mobile}.
Dependency-aware loaders treat browser pipelines as systems to measure and
optimize~\citep{netravali2016polaris,ruamviboonsuk2017vroom}.
High-fidelity crawling workloads similarly expose browser compute cost at
scale~\citep{goel2024sprinter}.
Website-complexity studies quantify how page structure drives cost~\citep{butkiewicz2011understanding}.
Closer to JSS concerns, empirical work on energy and performance practices shows that
``best practices'' must be validated, not assumed~\citep{procaccianti2016jss,malavolta2020webenergy}.
Selakovic and Pradel show that many JavaScript ``optimizations'' are engine-dependent
and do not transfer universally~\citep{selakovic2016jsperf}---a caution our
layer-translation results echo.

\subsection{Web Workers and OffscreenCanvas}
Web Workers move script off the UI thread~\citep{whatwg2024workers}.
OffscreenCanvas enables canvas contexts in workers~\citep{surma2018offscreencanvas,mdn2024offscreencanvas};
Smits studies the pattern in visualization systems~\citep{smits2020performance}.
These sources establish \emph{how} to offload; they do not attribute Worker versus
renderer versus sim backend on one DOM-sourced particle effect family.

\subsection{WebAssembly}
WebAssembly is a portable compilation target~\citep{haas2017wasm} with formal
foundations~\citep{watt2019mechanising}.
Empirical studies report WASM speedups over JavaScript on numerical workloads, with
gaps to native code~\citep{herrera2018numerical,jangda2019notsofast}, and document
runtime/energy differences across browsers~\citep{macedo2022wasm,vanhasselt2022wasmenergy,pockstaller2023wasmenergy}.
Dynamic analyses of JavaScript behaviour motivate why typed kernels can help~\citep{richards2010analysis}.
Most of this literature stops at the kernel or microbenchmark; we pair a sim-only
kernel bench with Worker+WebGL end-to-end FPS to show when kernel wins fail to
translate.

\subsection{WebGL, Canvas2D, and particle systems}
Canvas2D and WebGL2~\citep{webglSpec} trade convenience for GPU throughput.
Public particle demos~\citep{twojs2014particles} and theses comparing WebGPU/WebGL
particle paths~\citep{kronander2024webgpu} motivate GPU rendering at high~$N$.
None of these provide a controlled P1--P5-style decomposition that also varies thread
and WASM on a DOM-captured effect.

\subsection{DOM capture and non-particle dissolves}
DOM-to-canvas capture is common in practice~\citep{html2canvas}.
Multi-canvas CSS dissolves~\citep{thanosCssPractice} inspire our \emph{non-particle}
baseline P0, kept outside the P1--P5 particle decomposition.

\subsection{Jank instrumentation (background)}
RAIL~\citep{rail2015}, Chromium jank notes~\citep{chromiumJank}, Long
Tasks~\citep{longtasks}, LoAF~\citep{loaf2024}, and INP~\citep{inp2024} motivate
why main-thread congestion matters.
We report lab effect FPS/p95 as primary metrics.

\subsection{Positioning and research gap}
The scientific gap is not ``nobody has used Workers, WebGL, or WASM.''
Those technologies are mature and well studied in isolation
(Table~\ref{tab:rw-compare}).
The gap is \emph{claim hygiene for layered browser architectures}: prior empirical work
rarely holds the product workload fixed while varying thread, renderer, and simulation
backend together, and rarely pairs a kernel microbenchmark with end-to-end FPS so that
non-translation of a layer win can be observed rather than assumed away.
DOM-sourced particle pipelines are our instrumented instance of that broader problem;
particle aesthetics are incidental.

To the best of our knowledge, prior work has not provided a controlled, same-effect
decomposition across Worker $\times$ WebGL/Canvas2D $\times$ WASM with explicit
end-to-end and kernel regimes that demonstrate when a layer-level optimization fails to
propagate to user-visible performance.
After Table~\ref{tab:rw-compare}, a reviewer should be able to restate the gap in one
sentence: this paper supplies bottleneck-aware systems measurement for a multi-lever
browser pipeline, not another single-technology speedup report.

\begin{table}[t]
\centering
\caption{Coverage of related empirical/systems studies versus this work
(structured comparison, not an SLR).
Columns mark whether a study jointly varies or reports the corresponding concern.}
\label{tab:rw-compare}
\scriptsize
\begin{tabularx}{\linewidth}{Yccccccc}
\toprule
Study & Worker & WebGL & WASM & DOM & E2E & Kernel & Axes \\
\midrule
WProf / mobile browser
  \citep{wang2013wprof,nejati2016mobile}
  & -- & -- & -- & page & yes & -- & -- \\
JS perf.\ issues
  \citep{selakovic2016jsperf}
  & -- & -- & -- & -- & app & -- & -- \\
WASM design/emp.\
  \citep{haas2017wasm,jangda2019notsofast,herrera2018numerical,macedo2022wasm}
  & -- & -- & yes & -- & part. & yes & -- \\
OffscreenCanvas / viz
  \citep{smits2020performance}
  & yes & part. & -- & -- & yes & -- & -- \\
Web particles
  \citep{kronander2024webgpu}
  & -- & yes & -- & -- & yes & -- & -- \\
\textbf{This work}
  & yes & yes & yes & yes & yes & yes & yes \\
\bottomrule
\end{tabularx}
\end{table}

\section{Method: Pipeline Architectures}
\label{sec:method}

\subsection{Shared capture}
All particle pipelines start from the same capture of a target DOM box into an
\texttt{ImageBitmap} / pixel buffer, under shared geometry and styling constraints
(Fig.~\ref{fig:architecture}).
Capture cost is recorded as \texttt{captureMs} and feeds RQ3.
P0 is the exception: it splits the captured image across CSS layers rather than
sampling particles.

\subsection{Design axes}
We treat performance as a product of three orthogonal choices
(Table~\ref{tab:axes}):
\begin{itemize}
\item \textbf{Thread:} main UI thread vs dedicated Web Worker (OffscreenCanvas).
\item \textbf{Simulation:} JavaScript typed-array loop, AssemblyScript/WASM, or none
(CSS-layer motion).
\item \textbf{Renderer:} Canvas2D, WebGL2 instancing, or multi-canvas CSS layers.
\end{itemize}

\begin{table}[t]
\centering
\caption{Design axes varied across pipelines.}
\label{tab:axes}
\begin{tabularx}{\linewidth}{lY}
\toprule
Axis & Options \\
\midrule
Thread & \texttt{main}, \texttt{worker} \\
Simulation & \texttt{js}, \texttt{wasm}, \texttt{none} (CSS layers) \\
Renderer & \texttt{canvas2d}, \texttt{webgl2}, \texttt{css-layers} \\
\bottomrule
\end{tabularx}
\end{table}

\subsection{Six pipelines}
Table~\ref{tab:pipelines} lists the concrete configurations.
P1 is the negative control (particles on the main thread with Canvas2D).
P2 keeps simulation on the main thread but switches to WebGL2.
P3--P5 move work into a worker; among them, P4 and P5 share WebGL2 while differing
only in the simulation backend (JS vs WASM).
P5 mirrors our reference implementation (\texttt{React-Thanos-Effect}): OffscreenCanvas
WebGL2 in a worker, with the particle update written in AssemblyScript.

\begin{table}[t]
\centering
\caption{Pipelines: P1--P5 are the main particle decomposition; P0 is an additional
non-particle baseline.}
\label{tab:pipelines}
\small
\begin{tabularx}{\linewidth}{llllY}
\toprule
ID & Thread & Sim & Render & Role \\
\midrule
P0 & main & none & CSS (16) & Non-particle baseline \\
P1 & main & JS & Canvas2D & Negative control \\
P2 & main & JS & WebGL2 & GPU without offload \\
P3 & worker & JS & Canvas2D & Worker + Canvas2D \\
P4 & worker & JS & WebGL2 & Worker + GPU, no WASM \\
P5 & worker & WASM & WebGL2 & Full proposed stack \\
\bottomrule
\end{tabularx}
\end{table}

\paragraph{Main decomposition versus P0}
The controlled architectural decomposition that answers RQ1--RQ4 is \textbf{P1--P5}:
five particle pipelines that share DOM capture and particle sampling while varying
thread, simulation backend, and renderer (Fig.~\ref{fig:bottleneck}).
P0 is an \emph{additional non-particle baseline} (16 CSS layers) included for
paradigm contrast; it is not a per-particle peer and is never used to claim Worker /
WebGL / WASM axis effects.

Wave-mode motion is shared between the JS and WASM update entry points so the
backends stay comparable.
Particle budgets follow the same area/density formula; stress and sim-only raise the
cap to 250k--500k.

\subsection{Metrics}
Table~\ref{tab:metrics} summarizes the primary metrics.
Headed UI-sampler / long-task signals proved unreliable in our setup, so primary claims
use effect FPS and frame-time p95.
For sim-only we report mean update-frame time and derived particles/s, with no draw.

\begin{table}[t]
\centering
\caption{Primary metrics reported by the harness.}
\label{tab:metrics}
\begin{tabular}{lp{7.6cm}}
\toprule
Metric & Meaning \\
\midrule
\texttt{avgFps} & Mean effect FPS over the animation window \\
\texttt{frameTimeP95Ms} & 95th percentile inter-frame time of the effect \\
\texttt{captureMs} & Time to obtain the bitmap from the DOM target \\
\texttt{peakParticles} & Peak live particle count \\
\texttt{avgFrameMs} & Sim-only: mean update-kernel time (no draw) \\
\texttt{phases.*} & Optional P4/P5 in-worker means: simulate / viewPrep / upload / draw \\
\bottomrule
\end{tabular}
\end{table}

\section{Experimental Setup}
\label{sec:setup}

\subsection{Machine and browser}
Paper-primary rankings come from headed GPU runs on a laptop configuration
(Table~\ref{tab:machine}).
Same-host Firefox/Intel and Chrome/NVK slices use that laptop; the second-host
slice uses a Google Colab Tesla~T4 runtime (Table~\ref{tab:machine-colab}).
A headless SwiftShader session exists only as a secondary footnote and is not used for
WebGL ranking.

\begin{table}[t]
\centering
\caption{Paper-primary evaluation machine and browser.}
\label{tab:machine}
\small
\begin{tabularx}{\linewidth}{lY}
\toprule
Item & Value \\
\midrule
OS & Ubuntu 24.04 LTS \\
CPU & Intel Core i7-11800H (8C/16T) \\
RAM & 16\,GiB \\
GPU & NVIDIA GeForce RTX 3050 Mobile \\
Browser & Chrome 138 (Playwright \texttt{channel: chrome}, headed) \\
WebGL path & ANGLE + Vulkan / Mesa NVK (RTX 3050 Mobile) \\
Repetitions & $n{=}10$ per cell (primary + Firefox replication) \\
Library git & \texttt{99b72cd} (\texttt{React-Thanos-Effect}) \\
\bottomrule
\end{tabularx}
\end{table}

\begin{table}[t]
\centering
\caption{Second-host cloud configuration (Track~B replication).}
\label{tab:machine-colab}
\small
\begin{tabularx}{\linewidth}{lY}
\toprule
Item & Value \\
\midrule
Host & Cloud GPU runtime (Tesla T4) \\
OS & Ubuntu 22.04 \\
GPU & NVIDIA Tesla T4 \\
Browser & Chrome 150 (Playwright \texttt{channel: chrome}) \\
Display & Xvfb $1920{\times}1080$; Chrome headed (\texttt{HEADED=1}) \\
WebGL path & ANGLE + OpenGL (\texttt{CHROME\_GL=gl});
  NVIDIA userspace GL libs \\
WebGL renderer & ANGLE (NVIDIA, Tesla T4/PCIe/SSE2, OpenGL 4.5.0) \\
Repetitions & $n{=}5$ per cell (\texttt{bench:replicate}) \\
\bottomrule
\end{tabularx}
\end{table}

\subsection{Fixtures and box sizes}
\label{sec:fixtures}
We use three \emph{synthetic} UI fixtures that we author and control in the harness
(Table~\ref{tab:fixtures}).
They are not scraped pages, production templates, or a sample from the live web.
``DOM-sourced'' in this paper means only that particles are generated from a
\texttt{html2canvas}-style capture of a live DOM subtree in the page under test---i.e.,
the \emph{capture stage} is genuine---not that the fixtures represent a corpus of
real-world websites.

We chose synthetics deliberately for internal validity: shared geometry, stable
markup, and a controlled complexity ladder (A~$<$~B~$<$~C) that isolates capture cost
(RQ3) and keeps the architectural axes (thread / renderer / sim) comparable across
pipelines.
What we claim to generalize is the \emph{layer-translation} pattern under these
controlled DOM loads, not absolute FPS on arbitrary production DOMs
(CSS frameworks, deeply nested layouts, webfonts, iframes, etc.).
Interactive runs use a $600\times800$ CSS-pixel box; stress runs enlarge the box to
$1200\times1600$ to push particle counts and fill rate.

\begin{table}[t]
\centering
\caption{Synthetic DOM fixtures (harness-authored; not a web corpus).}
\label{tab:fixtures}
\small
\begin{tabularx}{\linewidth}{llY}
\toprule
ID & Content & Purpose \\
\midrule
A & Image-only & Light capture; pipeline ceiling \\
B & Simple card & Modest composite UI card \\
C & Heavy DOM ($8\times4$ grid) & Capture-pressure ladder step \\
\bottomrule
\end{tabularx}
\end{table}

\subsection{Regimes and matrices}
We deliberately split measurement into three regimes that answer different questions
(Table~\ref{tab:regimes}).
\begin{itemize}
\item \textbf{Interactive (official).}
Fixtures A,B $\times$ densities $\{0.5,1,2\}$ $\times$
\{P0 baseline, P1--P5\} $\times$ 10 reps,
plus C $\times$ density~1 $\times$ \{P1,P5\} $\times$ 10 $\rightarrow$
\textbf{380/380} formal OK.
Duration is about 1000\,ms of wave motion; peaks reach about 96k particles at
density~2 on the $600\times800$ box.
\item \textbf{Stress (end-to-end).}
Fixture~A, $1200\times1600$, densities $\{1,2,4\}$, pipelines \{P3,P4,P5\},
10 reps $\rightarrow$ \textbf{90/90} OK.
Peaks land near 192k (d=1) and 250k (d=2/4, capped).
\item \textbf{Sim-only.}
Backends \{\texttt{js},\texttt{wasm}\},
$N\in\{50\mathrm{k},100\mathrm{k},200\mathrm{k},250\mathrm{k},500\mathrm{k}\}$,
10 reps, 20 warmup + 200 timed frames $\rightarrow$ \textbf{100/100} OK.
Opaque $256\times256$ initialization; no draw.
\end{itemize}

\begin{table}[t]
\centering
\caption{Measurement regimes and what each isolates.}
\label{tab:regimes}
\begin{tabular}{llp{5.4cm}}
\toprule
Regime & Scale & Isolates \\
\midrule
Interactive & $600\times800$, $\lesssim$30k particles & Everyday UI pacing \\
Stress & $1200\times1600$, up to 250k & End-to-end FPS ceiling \\
Sim-only & Update only, $N$ to 500k & CPU sim throughput \\
\bottomrule
\end{tabular}
\end{table}

\subsection{Reproduction}
Library commit \texttt{99b72cd}
(\href{https://github.com/Hossein-Asadi/React-Thanos-Effect}{React-Thanos-Effect}).

Harness: \texttt{experiments/harness}; default \texttt{BENCH\_REPS=10}.
\begin{flushleft}\ttfamily\footnotesize
cd experiments/harness\\
HEADED=1 BENCH\_REPS=10 npm run bench:official\\
HEADED=1 BENCH\_REPS=10 npm run bench:stress\\
HEADED=1 BENCH\_REPS=10 npm run bench:sim\\
npx playwright install firefox\\
HEADED=1 BROWSER=firefox npm run bench:\{official,stress,sim\}
\end{flushleft}
Paper-primary trees: \texttt{experiments/results/chrome-138/}.
Firefox replication: \texttt{experiments/results/firefox-153/}.

\section{Results}
\label{sec:results}

Unless noted, interactive aggregates are means over ten repetitions; stress and
sim-only aggregates are medians over ten repetitions.

\subsection{Interactive regime}
\label{sec:res-interactive}

\begin{figure}[t]
\centering
\includegraphics[width=0.92\linewidth]{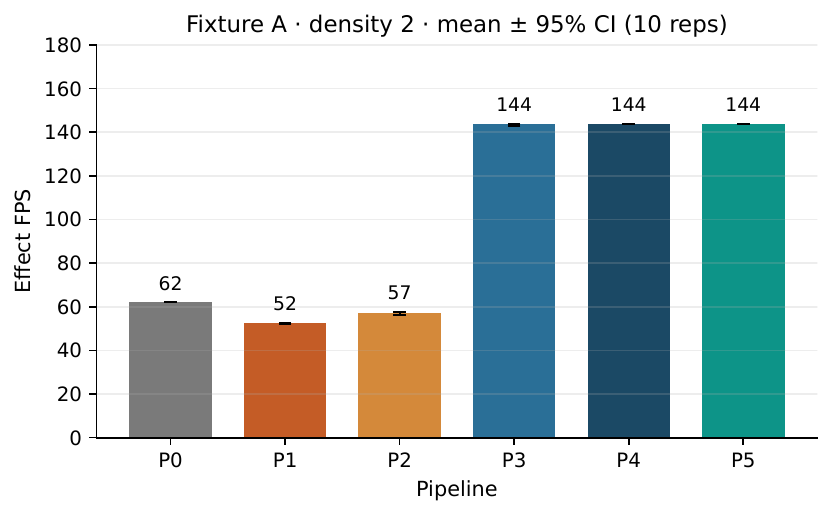}
\caption{Effect FPS on Fixture~A at density~2 (mean $\pm$ bootstrap 95\% CI, 10 reps).
Worker pipelines sit near a $\sim$144\,FPS ceiling; main-thread particle pipelines
(P1--P2) remain near 52--57.}
\label{fig:fps}
\end{figure}

Table~\ref{tab:fps-A} and Fig.~\ref{fig:fps} show effect FPS on Fixture~A.
Worker pipelines P3--P5 sit on a $\sim$144\,FPS ceiling at every density, while
main-thread particle pipelines remain near 52--60\,FPS.
P0 (non-particle CSS baseline) stays near 62\,FPS and is density-insensitive.
Putting WebGL on the main thread (P2) does not get into the worker class: at density~2,
P2 averages 57.0\,FPS versus 52.5\,FPS for P1.
Table~\ref{tab:stats-key} reports bootstrap 95\% CIs and Mann--Whitney tests for the
key comparisons; worker-versus-P1 separations are large, with complete sample
separation in the ten-run comparison (Cliff's $\delta{=}1.0$; exact
$p{=}1.08\times10^{-5}$), while P4-versus-P5 interactive FPS remains ceiling-tied
($p{=}0.74$).

\begin{table}[t]
\centering
\caption{Interactive effect FPS on Fixture~A (means over 10 reps).}
\label{tab:fps-A}
\begin{tabular}{lrrr}
\toprule
Pipeline & $d{=}0.5$ & $d{=}1$ & $d{=}2$ \\
\midrule
P0 CSS baseline & 63.4 & 62.0 & 62.0 \\
P1 Canvas2D main & 57.3 & 56.5 & 52.5 \\
P2 WebGL main & 59.8 & 60.0 & 57.0 \\
P3 Canvas2D worker & 143.9 & 143.9 & 143.5 \\
P4 WebGL worker (JS) & 143.9 & 143.8 & 143.9 \\
P5 Worker+WASM+WebGL & 143.9 & 143.9 & 143.9 \\
\bottomrule
\end{tabular}
\end{table}

\begin{table}[t]
\centering
\caption{Key paper-primary comparisons (Fixture~A, $n{=}10$).
Means with bootstrap 95\% CI; Mann--Whitney $U$ (exact, two-sided).
Cliff's $\delta{=}1$ indicates complete sample separation.
Raw trees under \texttt{experiments/results/chrome-138/}.}
\label{tab:stats-key}
\scriptsize
\begin{tabularx}{\linewidth}{Yrrrr}
\toprule
Comparison & Metric & Mean $\Delta$ & Cliff $\delta$ & $p$ \\
\midrule
P1 vs P3 & FPS (d2) & $+91.1$ & $1.00$ & $<10^{-4}$ \\
P1 vs P5 & FPS (d2) & $+91.4$ & $1.00$ & $<10^{-4}$ \\
P4 vs P5 & FPS (d2) & $+0.0$ & $0.10$ & $0.74$ \\
P4 vs P5 & stress FPS (d2) & $-10.0$ & $-0.35$ & $0.22$ \\
P3 vs P4 & stress FPS (d2) & $+42.8$ & $1.00$ & $<10^{-4}$ \\
JS vs WASM & sim ms @250k & $-0.92$ & $-1.00$ & $<10^{-4}$ \\
\bottomrule
\end{tabularx}
\end{table}

\begin{figure}[t]
\centering
\includegraphics[width=0.92\linewidth]{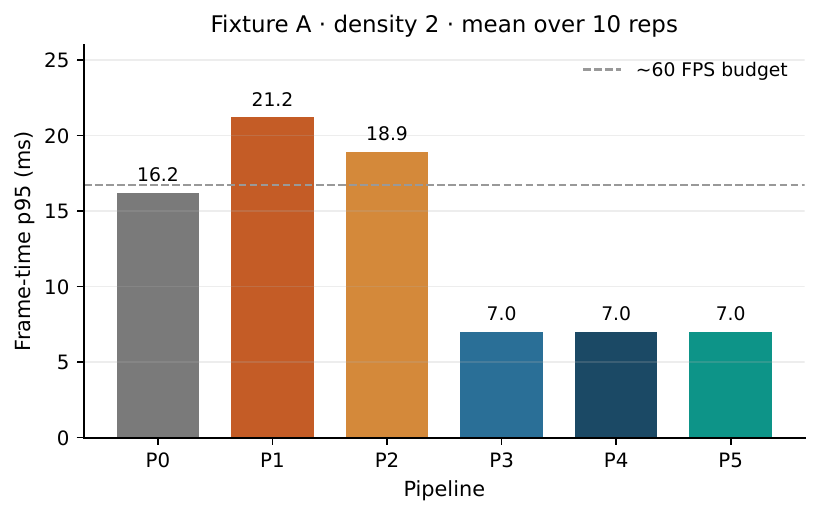}
\caption{Frame-time p95 on Fixture~A at density~2.
Workers improve pacing ($\approx$7\,ms) versus main-thread particle pipelines
($\approx$19--21\,ms).}
\label{fig:p95}
\end{figure}

Frame-time p95 tells the same story (Table~\ref{tab:p95-A}, Fig.~\ref{fig:p95}).
At density~2, workers land at 7.0\,ms while P1 sits at 21.2\,ms.
That is the interactive answer to RQ1: worker-class pipelines clearly beat the
main-thread Canvas2D particle baseline on both FPS and pacing.
The observed improvement is attributable primarily to worker offload; the interactive
ceiling prevents separating WebGL and WASM effects in this regime.

\begin{table}[t]
\centering
\caption{Interactive frame-time p95 and peak particles on Fixture~A at density~2.}
\label{tab:p95-A}
\setlength{\tabcolsep}{5pt}
\begin{tabular}{lrrr}
\toprule
Pipeline & p95 (ms) & avg frame (ms) & peak particles \\
\midrule
P0 CSS layers & 16.2 & 16.1 & --- \\
P1 Canvas2D main & 21.2 & 19.1 & 96{,}000 \\
P2 WebGL main & 18.9 & 17.5 & 96{,}000 \\
P3 Canvas2D worker & 7.0 & 7.0 & 96{,}000 \\
P4 WebGL worker & 7.0 & 7.0 & 96{,}000 \\
P5 Worker+WASM+WebGL & 7.0 & 7.0 & 96{,}000 \\
\bottomrule
\end{tabular}
\end{table}

\begin{figure}[t]
\centering
\includegraphics[width=0.92\linewidth]{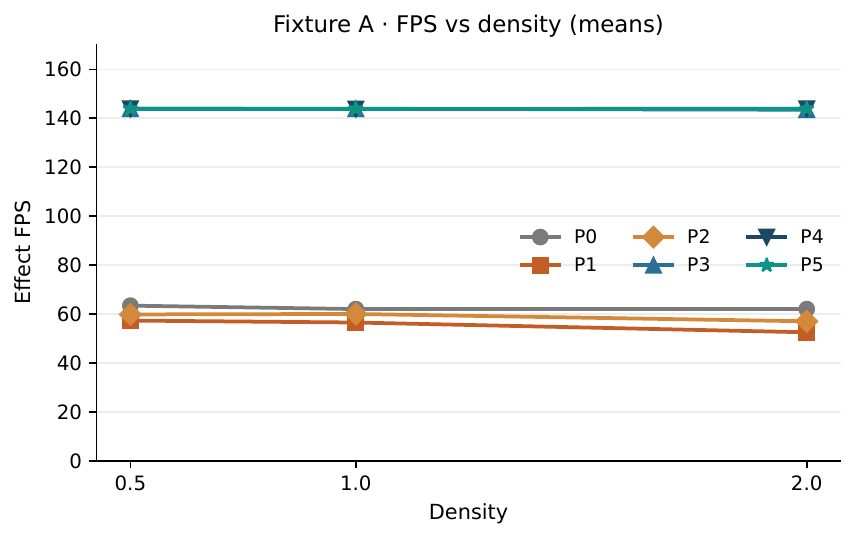}
\caption{Fixture~A FPS versus density.
Worker pipelines remain pinned near 144\,FPS across the interactive ladder.}
\label{fig:fps-density}
\end{figure}

Fig.~\ref{fig:fps-density} plots FPS against density.
Raising density from 0.5 to 2 does not separate P3/P4/P5 under the ceiling, so the
P5$-$P1 gap stays large ($\approx$91\,FPS at density~2).
For interactive RQ2, the density ladder therefore does not reopen gaps among workers;
the worker class is already saturated.

Fixture~B reproduces the same worker/main split at density~2
(Table~\ref{tab:fps-B}): workers again sit at $\approx$144\,FPS with p95
$\approx$7\,ms, while P1 drops to 51.9\,FPS with p95 22.1\,ms.
For interactive RQ4, the dominant axis is therefore \textbf{Worker offload}; WebGL and
WASM are not separable while the FPS ceiling holds.

\begin{table}[t]
\centering
\caption{Interactive summary on Fixture~B at density~2 (means, $n{=}10$).}
\label{tab:fps-B}
\begin{tabular}{lrr}
\toprule
Pipeline & FPS & p95 (ms) \\
\midrule
P0 CSS layers & 62.1 & 16.2 \\
P1 Canvas2D main & 51.9 & 22.1 \\
P2 WebGL main & 52.7 & 20.8 \\
P3 Canvas2D worker & 143.7 & 7.0 \\
P4 WebGL worker & 143.9 & 7.0 \\
P5 Worker+WASM+WebGL & 143.9 & 7.0 \\
\bottomrule
\end{tabular}
\end{table}

\subsection{Capture versus DOM complexity}
\label{sec:res-capture}

\begin{figure}[t]
\centering
\includegraphics[width=0.88\linewidth]{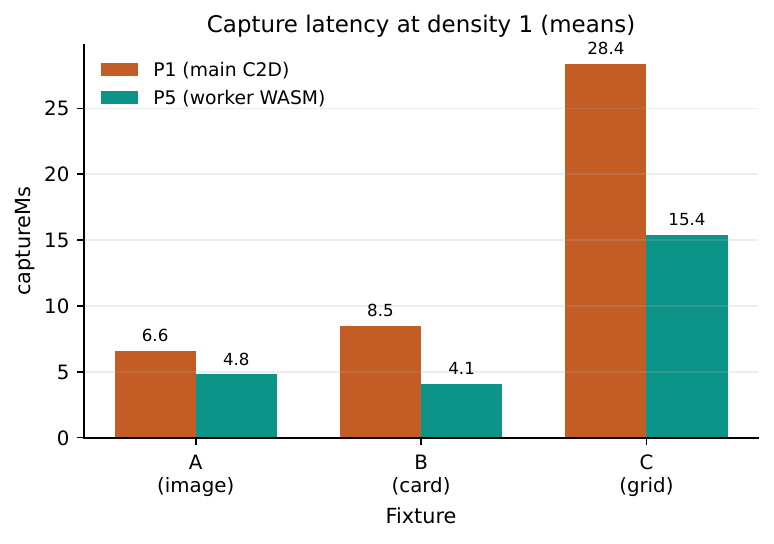}
\caption{Capture latency at density~1 for P1 and P5 ($n{=}10$).
P1 grows A~$<$~B~$<$~C; P5 is dominated by the heavy-grid fixture.}
\label{fig:capture}
\end{figure}

Capture is a first-class startup cost for DOM-sourced effects, not a detail that
disappears once animation starts.
Table~\ref{tab:capture} and Fig.~\ref{fig:capture} report \texttt{captureMs} at
density~1 ($n{=}10$).
For P1 the order is stable: A~$6.6$ $<$ B~$8.5$ $<$ C~$28.4$\,ms.
For P5, fixture~C remains clearly heaviest (15.4\,ms) while A/B stay in a light band
(4.8 / 4.1\,ms); the important RQ3 signal is that capture grows with DOM complexity and
is independent of the animate/render loop.
RQ3 holds.

\begin{table}[t]
\centering
\caption{Capture latency at density~1 (means, ms; $n{=}10$).}
\label{tab:capture}
\begin{tabular}{lrr}
\toprule
Fixture & P1 \texttt{captureMs} & P5 \texttt{captureMs} \\
\midrule
A (image-only) & 6.6 & 4.8 \\
B (simple card) & 8.5 & 4.1 \\
C (heavy grid) & 28.4 & 15.4 \\
\bottomrule
\end{tabular}
\end{table}

\subsection{Stress end-to-end}
\label{sec:res-stress}

\begin{figure}[t]
\centering
\includegraphics[width=0.92\linewidth]{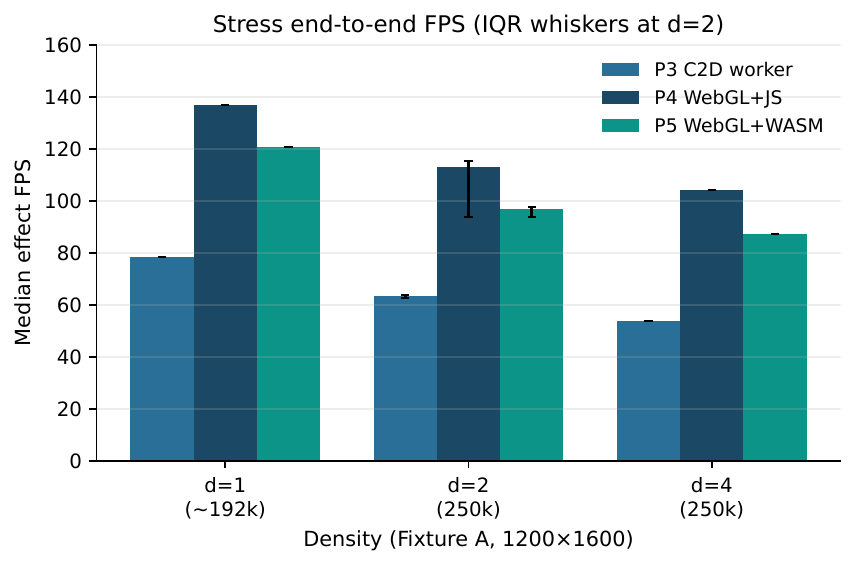}
\caption{Stress end-to-end median FPS ($n{=}10$, Chrome/NVK).
P4 leads; P5 stays below P4 under a shared WebGL-heavy regime.}
\label{fig:stress}
\end{figure}

\begin{figure}[t]
\centering
\includegraphics[width=0.92\linewidth]{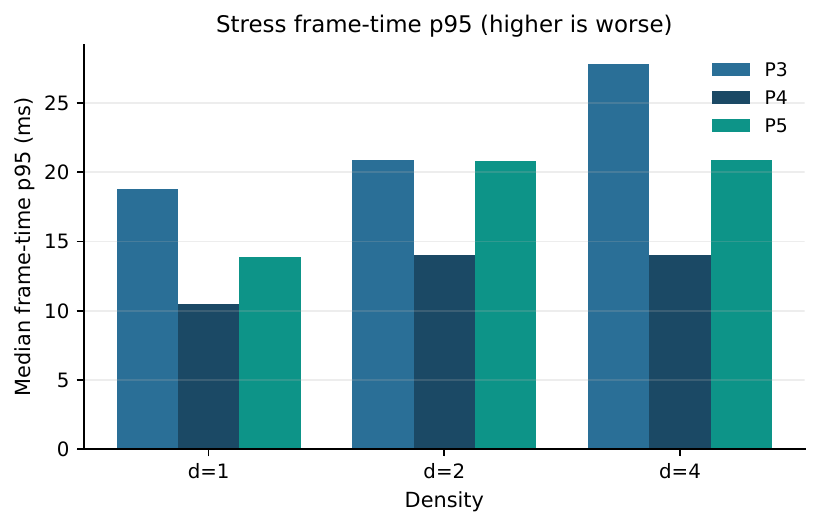}
\caption{Stress frame-time p95 ($n{=}10$).
P3 degrades most as density rises; WebGL workers keep lower p95 on this stack.}
\label{fig:stress-p95}
\end{figure}

The interactive ceiling disappears once we enlarge the box and raise particle counts.
Table~\ref{tab:stress} and Figs.~\ref{fig:stress}--\ref{fig:stress-p95} summarize
median FPS and p95 for worker pipelines under stress ($n{=}10$).
At density~2 (250k particles) on the paper-primary Chrome/NVK stack, P4 leads at
112.9\,FPS, ahead of P5 (96.9) and P3 (63.2).
Across densities, P5 stays below P4---there is no end-to-end WASM win---while WebGL
workers outperform the Canvas2D worker on this path.
An earlier headed Chrome stress matrix ($n{=}5$, archival) preferred P3 over WebGL,
so renderer ranking remains configuration-sensitive
(Section~\ref{sec:res-replicate}).
Density~4 matches density~2 in particle count because both hit the 250k cap.

\begin{table}[t]
\centering
\caption{Stress end-to-end medians on Fixture~A ($1200\times1600$; $n{=}10$).}
\label{tab:stress}
\begin{tabular}{lrrr}
\toprule
 & $d{=}1$ ($\sim$192k) & $d{=}2$ (250k) & $d{=}4$ (250k) \\
\midrule
\multicolumn{4}{l}{\emph{Median FPS}} \\
P3 Canvas2D worker & 78.4 & 63.2 & 53.7 \\
P4 WebGL+JS worker & \textbf{136.9} & \textbf{112.9} & \textbf{104.2} \\
P5 WebGL+WASM worker & 120.7 & 96.9 & 87.4 \\
\midrule
\multicolumn{4}{l}{\emph{Median frame-time p95 (ms)}} \\
P3 Canvas2D worker & 18.8 & 20.9 & 27.8 \\
P4 WebGL+JS worker & 10.5 & 14.0 & 14.0 \\
P5 WebGL+WASM worker & 13.9 & 20.8 & 20.9 \\
\bottomrule
\end{tabular}
\end{table}

Relative deltas make the claim discipline explicit (Table~\ref{tab:stress-delta}):
P5$-$P4 ranges from roughly $-12\%$ to $-16\%$ in median FPS, while P5 still beats P3
by about 50--60\% on this Chrome/NVK path.
For RQ2 under stress, gaps among workers reopen.
The portable negative result for RQ4 remains that the sim-only WASM win fails to raise
end-to-end FPS (P5$\le$P4 across configurations); Section~\ref{sec:res-phases}
reports where CPU time goes inside those P4/P5 frames.

\begin{table}[t]
\centering
\caption{Stress relative deltas (median FPS; $n{=}10$).}
\label{tab:stress-delta}
\begin{tabular}{lrr}
\toprule
Density & $\Delta$ P5 vs P4 & $\Delta$ P5 vs P3 \\
\midrule
1 & $-11.8\%$ & $+54.0\%$ \\
2 & $-14.1\%$ & $+53.3\%$ \\
4 & $-16.1\%$ & $+62.9\%$ \\
\bottomrule
\end{tabular}
\end{table}

\subsection{In-worker frame phases (P4/P5)}
\label{sec:res-phases}

To locate work inside stress frames, we instrumented the P4 and P5 worker loops with
per-frame \emph{CPU} timers for simulate, view prep, upload
(\texttt{bufferSubData}), and draw-call wall time, plus best-effort \emph{GPU}
elapsed time via WebGL2 disjoint timer queries wrapping the submitted
wipe, upload, and draw command stream.
We re-ran the stress matrix for those two methods on paper-primary Chrome/NVK
($n{=}10$; all cells succeeded).
The GPU extension is unavailable on our Firefox slice, so GPU means are Chrome-only.
Queries are asynchronous; end-of-run drainage uses \texttt{gl.finish}, and we discard
samples when \texttt{GPU\_DISJOINT\_EXT} is set (zero drops in this campaign).

\begin{figure}[t]
\centering
\includegraphics[width=0.92\linewidth]{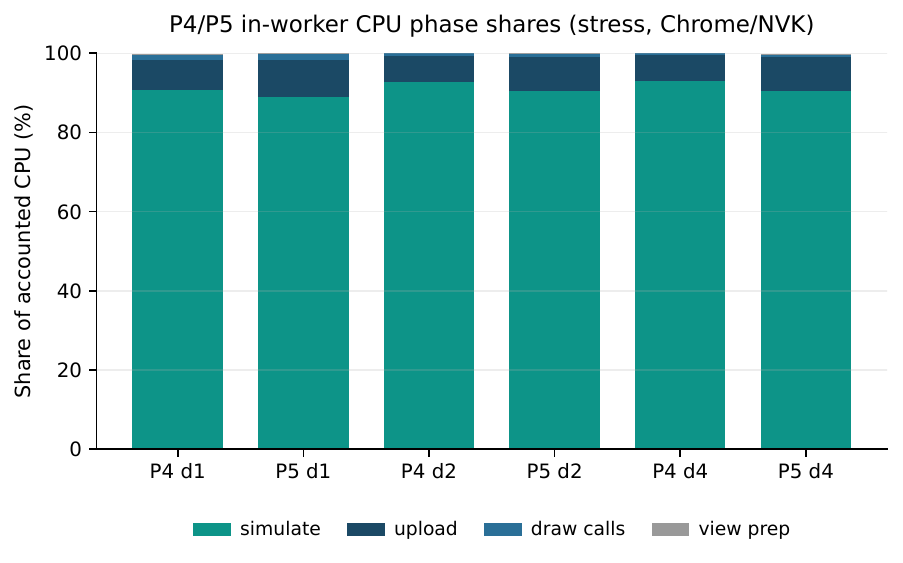}
\caption{Share of accounted in-worker CPU time under stress (Chrome/NVK).
Simulate dominates for both P4 and P5; upload is secondary; draw-call wall time is small.
GPU elapsed (Table~\ref{tab:phases}) is larger than drawCPU but still below simulate.}
\label{fig:phases}
\end{figure}

Table~\ref{tab:phases} and Fig.~\ref{fig:phases} summarize medians.
On the CPU side, simulate still occupies about $89$--$93\%$ of the accounted budget;
upload is next; draw-call wall time is about $1\%$.
GPU elapsed is far larger than that CPU draw wall (e.g., $\approx$2.2\,ms vs
$\approx$0.04\,ms at density~2), confirming that CPU-only draw timers understate GPU
work---yet GPU elapsed remains below the simulate CPU mean for both pipelines at every
tested density (P4 density~2: $2.19$ vs $5.45$\,ms; P5: $2.18$ vs $3.85$\,ms).
P4 and P5 GPU means are nearly identical at density~2--4, while P5's simulate CPU is
lower, consistent with the Chrome sim-only ordering.
Accounted CPU plus GPU timer still leaves slack relative to the inter-frame period at
observed stress FPS (scheduling/compositor/display outside these probes).
The portable negative result stands: the kernel win need not raise e2e FPS
(Table~\ref{tab:stress}); the instrumented path is not ``draw-call CPU dominated,''
and Chrome GPU timers do not show GPU wipe or draw exceeding simulate either.

\begin{table}[t]
\centering
\caption{Median in-worker phase means under stress (Fixture~A, $1200\times1600$;
Chrome/NVK; $n{=}10$).
\texttt{drawCPU} is CPU wall time around wipe+instanced draw;
\texttt{gpu} is mean \texttt{TIME\_ELAPSED} from disjoint timer queries.
sim\% is share of accounted CPU (simulate+view+upload+drawCPU).}
\label{tab:phases}
\begin{tabular}{llrrrrr}
\toprule
Method & $d$ & sim (ms) & upload (ms) & drawCPU (ms) & gpu (ms) & sim\% \\
\midrule
P4 & 1 & 2.88 & 0.25 & 0.04 & 0.85 & 90.7 \\
P5 & 1 & 2.17 & 0.23 & 0.04 & 1.20 & 88.9 \\
P4 & 2 & 5.45 & 0.40 & 0.04 & 2.19 & 92.7 \\
P5 & 2 & 3.85 & 0.37 & 0.04 & 2.18 & 90.5 \\
P4 & 4 & 6.68 & 0.48 & 0.04 & 2.34 & 93.0 \\
P5 & 4 & 5.21 & 0.49 & 0.04 & 2.31 & 90.5 \\
\bottomrule
\end{tabular}
\end{table}

\subsection{Sim-only microbench}
\label{sec:res-sim}

\begin{figure}[t]
\centering
\includegraphics[width=0.90\linewidth]{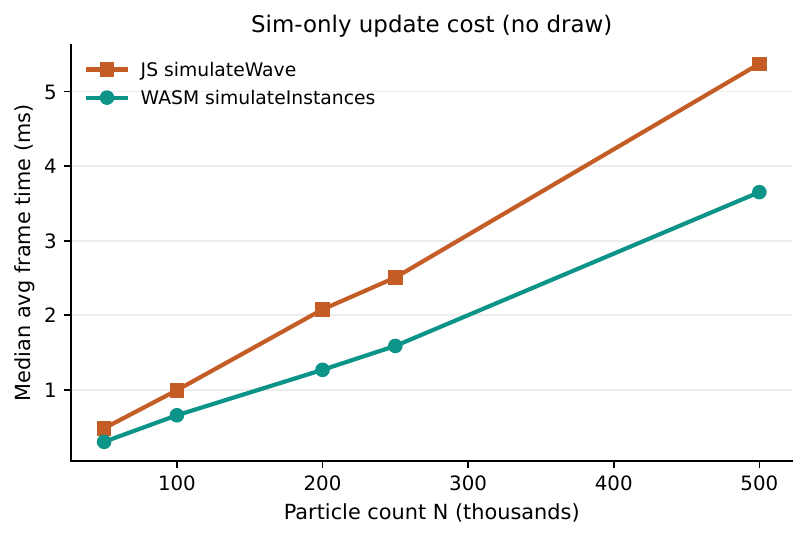}
\caption{Sim-only median frame time versus particle count.
WASM stays below the JS curve across the ladder.}
\label{fig:sim-ft}
\end{figure}

\begin{figure}[t]
\centering
\includegraphics[width=0.88\linewidth]{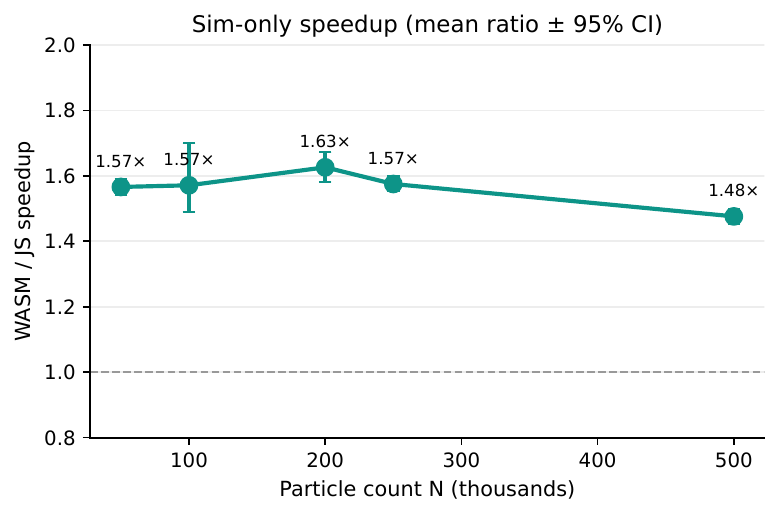}
\caption{Sim-only WASM/JS speedup (mean ratio with bootstrap 95\% CI), stable near
$1.5$--$1.6\times$ on paper-primary Chrome ($n{=}10$).}
\label{fig:sim}
\end{figure}

Isolating the update kernel flips the story again.
Table~\ref{tab:sim} and Figs.~\ref{fig:sim-ft}--\ref{fig:sim} show that WASM is
$1.47\times$--$1.64\times$ faster than the equivalent JS loop across
$N\in[5\cdot10^4,5\cdot10^5]$ ($n{=}10$).
Throughput stays in the $\sim$137--165M particles/s band for WASM versus
$\sim$93--103M for JS.
On the simulation kernel alone, the ordering is clear and consistent---this is the
supported half of the WASM claim in RQ4.

\begin{table}[t]
\centering
\caption{Sim-only median update cost (no draw; $n{=}10$).}
\label{tab:sim}
\begin{tabular}{rrrrr}
\toprule
$N$ & JS (ms) & WASM (ms) & JS part/s & Speedup \\
\midrule
50{,}000 & 0.484 & 0.303 & 103M & $1.60\times$ \\
100{,}000 & 0.995 & 0.659 & 100M & $1.51\times$ \\
200{,}000 & 2.078 & 1.268 & 96M & $1.64\times$ \\
250{,}000 & 2.506 & 1.589 & 100M & $1.58\times$ \\
500{,}000 & 5.372 & 3.651 & 93M & $1.47\times$ \\
\bottomrule
\end{tabular}
\end{table}

\subsection{Cross-browser and second-host replication}
\label{sec:res-replicate}

Paper-primary rankings use headed Chrome~138 with ANGLE/Vulkan (Mesa NVK) on the
evaluation laptop (Table~\ref{tab:machine}; $n{=}10$).
We probe external validity in two complementary ways.

\paragraph{Same-host cross-browser / dual-GPU-class}
We re-run the \emph{full} official (380), stress (90), and sim-only (100) matrices on
the same laptop under Firefox~153 with an Intel UHD WebGL renderer ($n{=}10$), and a
headed Chrome/NVK replication slice.

\paragraph{Independent second host (Tesla T4)}
We additionally run the minimum \texttt{bench:replicate} matrix
(interactive A/d2; stress A/d2 at $1200{\times}1600$;
sim-only $N\in\{100\mathrm{k},250\mathrm{k}\}$)
on a cloud Tesla~T4 VM with Chrome~150
(Table~\ref{tab:machine-colab}; $n{=}5$).
That host exposes CUDA and NVIDIA userspace GL libraries but does not provide a
working headless Vulkan path for Chrome; we therefore drive headed Chrome under Xvfb
with \texttt{CHROME\_GL=gl} and NVIDIA GL on the library path.
The recorded WebGL renderer is a hardware Tesla~T4 string (not SwiftShader).

Table~\ref{tab:replicate} reports the stress and sim-only outcomes that matter for
claim discipline (Fixture~A, density~2).
On the laptop stacks, under stress P5 stays below P4 and both prefer
P4${}>{}$P5${}>{}$P3.
On Colab T4 the stress ordering is P5$\approx$P4${}\gg{}$P3 (P5 slightly above P4 at
$n{=}5$; gap $\approx$2\,FPS), so we still reject a \emph{clear} end-to-end WASM win
while noting host-dependent absolute margins.
Chrome-class sim-only WASM/JS speedup holds on the primary host
($1.5$--$1.6\times$) and on Colab T4 ($\approx$1.85$\times$ at $N{=}250\mathrm{k}$),
while Firefox's sim-only WASM path is slower than JS
($\approx$0.56$\times$ at $N{=}250\mathrm{k}$).
Absolute interactive FPS also differ: Firefox workers are not pinned to the Chrome
$\sim$144\,FPS ceiling, and the Colab/Xvfb slice appears paced near $\sim$60\,FPS
(P4/P5 $\approx$60; P1 $\approx$54; P3 $\approx$45).
An archival headed Chrome stress matrix ($n{=}5$) preferred P3 over WebGL, so
renderer ranking can reverse across browser/GPU/driver/host configurations.
We therefore treat comparisons as rank and axis evidence rather than matched absolute
FPS: same-host multi-browser / dual-GPU-class replication plus one independent
cloud second host.

\begin{table}[t]
\centering
\caption{Stress/sim claims (Fixture~A, density~2; medians).
Laptop columns are full-matrix $n{=}10$; second host is \texttt{bench:replicate} $n{=}5$.
Archival $n{=}5$ Chrome stress preferred P3 (65.3 / 61.8 / 59.7).}
\label{tab:replicate}
\small
\begin{tabularx}{\linewidth}{Yrrr}
\toprule
Metric & Chrome/NVK & FF+Intel & 2nd host \\
\midrule
\multicolumn{4}{l}{\emph{Stress median FPS}} \\
P3 Canvas2D worker & 63.2 & 71.4 & 27.1 \\
P4 WebGL+JS worker & 112.9 & 164.9 & 57.0 \\
P5 WebGL+WASM worker & 96.9 & 102.3 & 58.2 \\
\midrule
\multicolumn{4}{l}{\emph{Sim-only WASM/JS @250k}} \\
Speedup ($t_{\mathrm{JS}}/t_{\mathrm{WASM}}$)
  & $1.58\times$ & $0.56\times$ & $1.85\times$ \\
\bottomrule
\end{tabularx}
\end{table}

\section{Discussion}
\label{sec:discussion}

\subsection{Answering the research questions}
Table~\ref{tab:rq} compresses the answers.

\begin{table}[t]
\centering
\caption{RQ answers grounded in the three regimes plus replication.}
\label{tab:rq}
\small
\begin{tabularx}{\linewidth}{lY}
\toprule
RQ & Answer \\
\midrule
RQ1 & Yes on paper-primary Chrome---primarily via worker offload (P3--P5 vs P1;
$n{=}10$); the interactive ceiling masks WebGL/WASM differences. \\
RQ2 & Interactive: flat at the Chrome ceiling when it binds. Stress: gaps reopen;
Chrome/NVK and Firefox prefer P4${}>{}$P5${}>{}$P3; Colab T4 has P5$\approx$P4${}\gg{}$P3
(archival Chrome preferred P3). \\
RQ3 & Capture grows with fixture complexity (P1: A$<$B$<$C); DOM capture is itself a
potential startup bottleneck. \\
RQ4 & Worker helps interactive regimes on paper-primary Chrome; WASM helps Chrome-class
sim-only (primary and second host); under high-load WebGL stress the kernel win fails to
translate into a clear e2e gain (P5$\le$P4 primary/Firefox; P5$\approx$P4 on second host)
even though CPU phase timers remain simulate-heavy (Section~\ref{sec:res-phases}). \\
\bottomrule
\end{tabularx}
\end{table}

\subsection{Finding / explanation / insight}
We structure the main lessons for JSS readers as finding, mechanism, and
generalizable insight (Fig.~\ref{fig:bottleneck}).

\paragraph{Finding 1 --- Worker improves interactive pacing}
On paper-primary Chrome ($n{=}10$), P3--P5 sit near $144$\,FPS while P1 stays near
$52$\,FPS (Table~\ref{tab:stats-key}; Cliff's $\delta{=}1$).
\emph{Explanation:} the interactive regime is limited by main-thread contention and
display/timer coupling; moving simulate+draw off the UI thread removes that bottleneck.
\emph{Insight:} when profiling shows UI-thread jank on an effect, offload first---before
reaching for WASM or a heavier renderer.

\paragraph{Finding 2 --- WASM improves the Chrome simulation kernel}
Sim-only AssemblyScript is $\sim$1.5--$1.6\times$ faster than the matched JS loop
($p{<}10^{-4}$ at $N{=}250\mathrm{k}$; speedup CI at $250\mathrm{k}$:
$[1.55,1.60]$).
\emph{Explanation:} the microbench isolates typed numeric updates without draw.
\emph{Insight:} a kernel speedup is real evidence about the simulation layer; it is not,
by itself, evidence about product FPS.
The Firefox replication demonstrates that this kernel advantage is not portable across
browser engines: sim-only WASM is slower than JS there
($\approx$0.56$\times$; Section~\ref{sec:res-replicate}).
The Colab T4 second host keeps a Chrome-class WASM win ($\approx$1.85$\times$ at
$N{=}250\mathrm{k}$).
Thus the speedup should be interpreted as a Chrome-class empirical result rather than a
general property of the WASM implementation across engines.

\paragraph{Finding 3 --- Kernel wins need not raise end-to-end FPS}
Under stress, P5 stays below P4 (Table~\ref{tab:stats-key}, $p{=}0.22$ at density~2;
Cliff's $\delta{=}-0.35$).
\emph{Explanation:} CPU phase timers show simulate still dominating the accounted
budget ($\approx$90\% at density~2); Chrome GPU timer queries show wipe, upload,
and draw GPU elapsed $\approx$2\,ms---much larger than draw-call CPU wall time, but still
below simulate for both P4 and P5, and nearly equal across the two pipelines
(Section~\ref{sec:res-phases}).
P5's simulate CPU is lower than P4's, yet primary e2e FPS still favors P4.
\emph{Insight:} microbenchmark optimization should not be interpreted as product-level
improvement without naming which layer---and which clock (CPU vs GPU)---was measured.

\paragraph{Finding 4 --- Renderer ranking can reverse}
The $n{=}10$ Chrome/NVK and Firefox matrices prefer P4${}>{}$P5${}>{}$P3 under stress;
Colab T4 prefers WebGL workers over Canvas2D (P5$\approx$P4${}\gg{}$P3); an archival
headed Chrome stress matrix ($n{=}5$) preferred P3
(Section~\ref{sec:res-replicate}).
\emph{Explanation:} render-path cost is GPU/driver/engine/host specific.
\emph{Insight:} do not crown Canvas2D or WebGL universally inside a worker; measure on
the target configuration.

\paragraph{Finding 5 --- Capture is an independent cost}
\texttt{captureMs} grows A${}<$B${}<$C for both P1 and P5 (RQ3).
\emph{Explanation:} fixture DOM complexity lengthens bitmap acquisition before any
particle frame runs.
\emph{Insight:} even with synthetic fixtures, DOM capture is a first-class startup
cost orthogonal to the animate/render loop; production pages would likely amplify
that cost, which is why we treat RQ3 as an independent bottleneck rather than as
evidence about the live web at large.

\subsection{Claim discipline}
Table~\ref{tab:claims} states what the data support and what they do not.
The practical takeaway is to name the bottleneck layer with the claim.

\begin{table}[t]
\centering
\caption{Claim checklist against paper-primary and replicate regimes.}
\label{tab:claims}
\small
\begin{tabularx}{\linewidth}{YYY}
\toprule
Claim & Evidence & Verdict \\
\midrule
Worker offload improves interactive FPS/p95 vs main-thread particles
  & Official A/B ($n{=}10$)
  & Supported (Chrome primary); host-dependent on second host/Xvfb \\
WASM update faster than equivalent JS
  & Chrome sim-only $1.5$--$1.6\times$; second host $1.85\times$
  & Supported on Chrome-class hosts; not on Firefox slice \\
WASM improves e2e FPS vs JS+WebGL at $\sim$250k
  & Stress P5$\le$P4 (laptop); P5$\approx$P4 (second host, $n{=}5$)
  & Not supported as a clear win \\
Canvas2D worker beats WebGL under stress
  & Archival Chrome P3$\ge$P4; $n{=}10$ stacks and second host prefer WebGL
  & Hardware-specific \\
\bottomrule
\end{tabularx}
\end{table}

\subsection{Implications for software and systems engineering}
These results matter beyond dissolve effects because they constrain how browser-facing
systems should be optimized, documented, and evaluated.

\begin{enumerate}
\item \textbf{Library and framework authors.}
Shipping a ``Worker + WebGL + WASM'' stack is not itself evidence of end-to-end
improvement.
Document which layer each release optimizes, and publish at least one product-FPS
regime alongside any kernel microbenchmark used in marketing or READMEs.
\item \textbf{Performance engineers and architects.}
Treat Worker, renderer choice, and WASM as orthogonal architectural decisions with
regime-dependent payoffs.
Profile for the dominant bottleneck first; offload before rewriting kernels when
main-thread contention dominates, and do not interpret a sim-only speedup as a
shipping FPS win when e2e FPS fails to move.
\item \textbf{Empirical evaluation and claim discipline.}
Browser performance claims should name the measurement regime (interactive, stress,
kernel-only), the engine/GPU configuration, and whether the observed win was
layer-local or end-to-end.
Cross-engine portability must be tested: our Firefox slice reverses the Chrome WASM
kernel ordering.
\item \textbf{Relevance to JSS-style software systems research.}
The transferable lesson is methodological---bottleneck-aware decomposition of layered
platform stacks---not a ranking of particle libraries.
The same validation pattern applies to other browser-mediated systems that compose
thread offload, compiled kernels, and GPU paths (visualization runtimes, creative
tools, and interactive dashboards).
\end{enumerate}

\section{Threats to Validity}
\label{sec:threats}

\textbf{External validity.}
Paper-primary and Firefox/NVK matrices use one laptop.
We add one \textbf{independent second host}: Google Colab Tesla~T4 with Chrome~150
under Xvfb (Section~\ref{sec:res-replicate}; $n{=}5$ replicate matrix).
That cloud slice corroborates Chrome-class sim-only WASM speedups and the absence of a
\emph{clear} e2e WASM stress win, but absolute interactive FPS are display-paced
($\sim$60\,Hz) and P3 underperforms P1 there---so we still treat absolute FPS as
host-specific.
Further hosts (Windows+proprietary NVIDIA, AMD, macOS, physical second laptops)
remain useful.

Fixtures are \emph{synthetic and harness-authored} (Section~\ref{sec:fixtures}).
They exercise a controlled capture-complexity ladder, but they are not a
representative sample of production websites.
Heavier real pages (large style recalculation, webfonts, nested compositing layers,
third-party widgets) may inflate \texttt{captureMs} and change how often the animate
loop shares the main thread with other work; our architectural rankings could then
shift in magnitude even if the layer-translation pattern remains useful.
We therefore do \emph{not} claim that absolute FPS, or the A/B/C capture latencies,
transfer to arbitrary live DOMs---only that, under these controlled fixtures, Worker /
WebGL / WASM show the reported regime-dependent gains and non-gains.
Evaluating the same axes on a curated corpus of real pages is future work.

\textbf{Construct validity.}
Headed \texttt{uiAvgFps} / long-task signals were unreliable, so we prioritize effect
FPS and p95.
Field metrics such as INP/LoAF remain useful complements but are not the primary tables.
Cross-engine absolute FPS are not directly comparable when display/timer ceilings differ.
Statistical tests on paper-primary tables use $n{=}10$ reps; the Colab second-host
slice uses $n{=}5$.
We still emphasize effect sizes and CIs alongside $p$-values
(Table~\ref{tab:stats-key}).

\textbf{Internal validity.}
Interactive worker pipelines on paper-primary Chrome hit a display/timer ceiling
($\sim$144\,FPS), which masks renderer and sim differences until stress.
P0 is a non-particle baseline and is not part of the P1--P5 axis decomposition.
Sim-only times the kernel without draw; the Chrome claim is the relative JS vs WASM
ordering.
P4/P5 CPU phase timers measure wall time around simulate/view/upload/draw.
GPU means use \texttt{EXT\_disjoint\_timer\_query\_webgl2} on Chrome only (absent on
our Firefox slice); they can be noisy and require end-of-run \texttt{gl.finish}, so
we treat them as supportive evidence rather than a portable primary metric.
Compositor/display cost outside the WebGL command stream remains unmeasured.
Stress densities 2 and 4 share the 250k particle cap.
WebGL renderer strings are logged with every result tree.

\section{Conclusion}
\label{sec:conclusion}

Using DOM-sourced particle pipelines as a controllable browser workload, we decomposed
a layered architecture along thread, renderer, and simulation axes (P1--P5, with P0 as
a non-particle baseline) to show when optimizing one layer fails to translate into
end-to-end gains.
Empirically: Worker offload improves interactive pacing on paper-primary Chrome;
WASM improves Chrome-class simulation kernels by about $1.5$--$1.85\times$ (not on our
Firefox slice); under WebGL-heavy load those kernel gains need not appear as a clear
end-to-end FPS win even when in-worker CPU phases remain simulate-dominated; and
renderer ranking can reverse across browser/GPU/host configurations.
A Colab Tesla~T4 second-host slice supports the Chrome-class kernel story and the
``no clear e2e WASM win'' reading under stress.
The practical lesson for software systems is not to select Worker, WebGL, or WASM as
universally faster technologies, but to identify the dominant bottleneck layer and
evaluate whether an optimization at that layer propagates to end-to-end user-visible
performance.

\paragraph{Future work}
Additional physical second hosts beyond Colab; compositor and display
instrumentation beyond WebGL timer queries; and product scenarios that remain
clearly sim-bound, where end-to-end WASM gains would be expected to reappear.

\section*{Availability}
\begin{sloppypar}
The reference library is at
\url{https://github.com/Hossein-Asadi/React-Thanos-Effect}
(commit \texttt{99b72cd}; 500k particle budget and optional P4/P5 CPU/GPU phase
timers on worker \texttt{complete}).
The measurement harness, fixtures, analysis notes, statistics scripts, and raw JSON
for the primary, same-host replicate, phase-instrumented stress, and second-host
regimes are in the companion \texttt{experiments/} package
(\texttt{harness/}, \texttt{results/}, \texttt{analysis/}), mirrored in that repository
and intended as supplementary material with this submission.
Some result records stamp \texttt{gitSha=605d2e7} (repository HEAD at collection
time); the measured library tree matches \texttt{99b72cd}.
Reproduce from \texttt{experiments/harness} with
\texttt{HEADED=1} and
\texttt{npm run bench:official},
\texttt{bench:stress},
\texttt{bench:sim}, or
\texttt{bench:replicate}
(\texttt{BROWSER=firefox} or \texttt{chrome} for replication;
for CPU+GPU phase stress set
\texttt{STRESS\_METHODS} to \texttt{P4-webgl-worker,P5-ours}).
This study uses synthetic UI fixtures only (no human subjects or personal data).
Competing interests: none declared.
Funding: this research did not receive a specific grant from funding agencies in the
public, commercial, or not-for-profit sectors.
A DOI-backed archive of the result trees, harness, and analysis scripts is at
\url{https://doi.org/10.5281/zenodo.21626647}.
\end{sloppypar}

\bibliographystyle{elsarticle-harv}
\bibliography{refs}

\begin{thebibliography}{31}
\expandafter\ifx\csname natexlab\endcsname\relax\def\natexlab#1{#1}\fi
\providecommand{\url}[1]{\texttt{#1}}
\providecommand{\href}[2]{#2}
\providecommand{\path}[1]{#1}
\providecommand{\DOIprefix}{doi:}
\providecommand{\ArXivprefix}{arXiv:}
\providecommand{\URLprefix}{URL: }
\providecommand{\Pubmedprefix}{pmid:}
\providecommand{\doi}[1]{\href{http://dx.doi.org/#1}{\path{#1}}}
\providecommand{\Pubmed}[1]{\href{pmid:#1}{\path{#1}}}
\providecommand{\bibinfo}[2]{#2}
\ifx\xfnm\relax \def\xfnm[#1]{\unskip,\space#1}\fi
\bibitem[{{Bluewings}(2019)}]{thanosCssPractice}
\bibinfo{author}{{Bluewings}}, \bibinfo{year}{2019}.
\newblock \bibinfo{title}{{Thanos} easter egg, how it works}.
\newblock \bibinfo{howpublished}{Technical reverse-engineering write-up}.
\newblock \URLprefix \url{https://bluewings.github.io/en/thanos-explained/}.
  \bibinfo{note}{documents multi-canvas + CSS dissolve pattern popularized by
  Google Search; accessed 2026-07-25}.
\bibitem[{Butkiewicz et~al.(2011)Butkiewicz, Madhyastha and
  Sekar}]{butkiewicz2011understanding}
\bibinfo{author}{Butkiewicz, M.}, \bibinfo{author}{Madhyastha, H.V.},
  \bibinfo{author}{Sekar, V.}, \bibinfo{year}{2011}.
\newblock \bibinfo{title}{Understanding website complexity: Measurements,
  metrics, and implications}, in: \bibinfo{booktitle}{Proceedings of the 2011
  ACM SIGCOMM Conference on Internet Measurement Conference},
  \bibinfo{publisher}{ACM}. pp. \bibinfo{pages}{313--328}.
\newblock \DOIprefix\doi{10.1145/2068816.2068846}.
\bibitem[{{Chrome Developers}(2024)}]{loaf2024}
\bibinfo{author}{{Chrome Developers}}, \bibinfo{year}{2024}.
\newblock \bibinfo{title}{Long animation frames {API}}.
\newblock \bibinfo{howpublished}{developer.chrome.com}.
\newblock \URLprefix
  \url{https://developer.chrome.com/docs/web-platform/long-animation-frames}.
  \bibinfo{note}{accessed 2026-07-25}.
\bibitem[{Goel et~al.(2024)Goel, Zhu, Netravali and
  Madhyastha}]{goel2024sprinter}
\bibinfo{author}{Goel, A.}, \bibinfo{author}{Zhu, J.},
  \bibinfo{author}{Netravali, R.}, \bibinfo{author}{Madhyastha, H.V.},
  \bibinfo{year}{2024}.
\newblock \bibinfo{title}{Sprinter: Speeding up high-fidelity crawling of the
  modern web}, in: \bibinfo{booktitle}{21st USENIX Symposium on Networked
  Systems Design and Implementation}, \bibinfo{publisher}{USENIX Association}.
\newblock \URLprefix
  \url{https://www.usenix.org/conference/nsdi24/presentation/goel}.
\bibitem[{{Google}(2015)}]{rail2015}
\bibinfo{author}{{Google}}, \bibinfo{year}{2015}.
\newblock \bibinfo{title}{{RAIL} model --- measure performance with the {RAIL}
  model}.
\newblock \bibinfo{howpublished}{web.dev}.
\newblock \URLprefix \url{https://web.dev/articles/rail}.
  \bibinfo{note}{accessed 2026-07-25}.
\bibitem[{Haas et~al.(2017)Haas, Rossberg, Schuff, Titzer, Gohman, Wagner,
  Zakai, Bastien and Holman}]{haas2017wasm}
\bibinfo{author}{Haas, A.}, \bibinfo{author}{Rossberg, A.},
  \bibinfo{author}{Schuff, D.L.}, \bibinfo{author}{Titzer, B.L.},
  \bibinfo{author}{Gohman, D.}, \bibinfo{author}{Wagner, L.},
  \bibinfo{author}{Zakai, A.}, \bibinfo{author}{Bastien, J.},
  \bibinfo{author}{Holman, M.}, \bibinfo{year}{2017}.
\newblock \bibinfo{title}{Bringing the web up to speed with {WebAssembly}}, in:
  \bibinfo{booktitle}{Proceedings of the 38th ACM SIGPLAN Conference on
  Programming Language Design and Implementation}, \bibinfo{publisher}{ACM}.
  pp. \bibinfo{pages}{185--200}.
\newblock \DOIprefix\doi{10.1145/3062341.3062363}.
\bibitem[{van Hasselt et~al.(2022)van Hasselt, Huijzendveld, Noort, de~Ruijter,
  Islam and Malavolta}]{vanhasselt2022wasmenergy}
\bibinfo{author}{van Hasselt, M.}, \bibinfo{author}{Huijzendveld, K.},
  \bibinfo{author}{Noort, N.}, \bibinfo{author}{de~Ruijter, S.},
  \bibinfo{author}{Islam, T.}, \bibinfo{author}{Malavolta, I.},
  \bibinfo{year}{2022}.
\newblock \bibinfo{title}{Comparing the energy efficiency of {WebAssembly} and
  {JavaScript} in web applications on {Android} mobile devices}, in:
  \bibinfo{booktitle}{Proceedings of the 26th International Conference on
  Evaluation and Assessment in Software Engineering}, \bibinfo{publisher}{ACM}.
\newblock \DOIprefix\doi{10.1145/3530019.3530035}.
\bibitem[{Herrera et~al.(2018)Herrera, Chen, Lavoie and
  Hendren}]{herrera2018numerical}
\bibinfo{author}{Herrera, D.}, \bibinfo{author}{Chen, H.},
  \bibinfo{author}{Lavoie, E.}, \bibinfo{author}{Hendren, L.},
  \bibinfo{year}{2018}.
\newblock \bibinfo{title}{Numerical computing on the web: Benchmarking for the
  future}, in: \bibinfo{booktitle}{Proceedings of the 14th ACM SIGPLAN
  International Symposium on Dynamic Languages}, \bibinfo{publisher}{ACM}. pp.
  \bibinfo{pages}{88--100}.
\newblock \DOIprefix\doi{10.1145/3276945.3276968}.
\bibitem[{von Hertzen and contributors(2024)}]{html2canvas}
\bibinfo{author}{von Hertzen, N.}, \bibinfo{author}{contributors},
  \bibinfo{year}{2024}.
\newblock \bibinfo{title}{{html2canvas}}.
\newblock \bibinfo{howpublished}{GitHub repository}.
\newblock \URLprefix \url{https://github.com/niklasvh/html2canvas}.
  \bibinfo{note}{accessed 2026-07-25}.
\bibitem[{Jangda et~al.(2019)Jangda, Powers, Berger and
  Guha}]{jangda2019notsofast}
\bibinfo{author}{Jangda, A.}, \bibinfo{author}{Powers, B.},
  \bibinfo{author}{Berger, E.D.}, \bibinfo{author}{Guha, A.},
  \bibinfo{year}{2019}.
\newblock \bibinfo{title}{Not so fast: Analyzing the performance of
  {WebAssembly} vs.\ native code}, in: \bibinfo{booktitle}{2019 USENIX Annual
  Technical Conference}, \bibinfo{publisher}{USENIX Association}. pp.
  \bibinfo{pages}{107--120}.
\newblock \URLprefix
  \url{https://www.usenix.org/conference/atc19/presentation/jangda}.
\bibitem[{{Khronos Group}(2022)}]{webglSpec}
\bibinfo{author}{{Khronos Group}}, \bibinfo{year}{2022}.
\newblock \bibinfo{title}{{WebGL} 2.0 specification}.
\newblock \bibinfo{howpublished}{Khronos}.
\newblock \URLprefix
  \url{https://registry.khronos.org/webgl/specs/latest/2.0/}.
  \bibinfo{note}{accessed 2026-07-25}.
\bibitem[{Kronander(2024)}]{kronander2024webgpu}
\bibinfo{author}{Kronander, J.}, \bibinfo{year}{2024}.
\newblock \bibinfo{title}{Performance Comparison of {WebGPU} and {WebGL} for
  {2D} Particle Systems on the Web}.
\newblock Master's thesis. Blekinge Institute of Technology.
\newblock \URLprefix
  \url{https://www.diva-portal.org/smash/get/diva2:1945245/FULLTEXT02}.
\bibitem[{de~Macedo et~al.(2022)de~Macedo, Abreu, Pereira and
  Saraiva}]{macedo2022wasm}
\bibinfo{author}{de~Macedo, J.}, \bibinfo{author}{Abreu, R.},
  \bibinfo{author}{Pereira, R.}, \bibinfo{author}{Saraiva, J.},
  \bibinfo{year}{2022}.
\newblock \bibinfo{title}{{WebAssembly} versus {JavaScript}: Energy and runtime
  performance}, in: \bibinfo{booktitle}{2022 International Conference on ICT
  for Sustainability}, \bibinfo{publisher}{IEEE}. pp. \bibinfo{pages}{24--34}.
\newblock \DOIprefix\doi{10.1109/ICT4S55073.2022.00014}.
\bibitem[{Malavolta et~al.(2020)Malavolta, Chinnappan, Jasmontas, Gupta and
  Soltany}]{malavolta2020webenergy}
\bibinfo{author}{Malavolta, I.}, \bibinfo{author}{Chinnappan, K.},
  \bibinfo{author}{Jasmontas, L.}, \bibinfo{author}{Gupta, S.},
  \bibinfo{author}{Soltany, K.A.K.}, \bibinfo{year}{2020}.
\newblock \bibinfo{title}{Evaluating the impact of caching on the energy
  consumption and performance of progressive web apps}, in:
  \bibinfo{booktitle}{Proceedings of the IEEE/ACM 7th International Conference
  on Mobile Software Engineering and Systems}, \bibinfo{publisher}{ACM}. pp.
  \bibinfo{pages}{11--20}.
\newblock \DOIprefix\doi{10.1145/3387905.3388593}.
\bibitem[{{MDN Contributors}(2024)}]{mdn2024offscreencanvas}
\bibinfo{author}{{MDN Contributors}}, \bibinfo{year}{2024}.
\newblock \bibinfo{title}{{OffscreenCanvas} --- {Web APIs}}.
\newblock \bibinfo{howpublished}{MDN Web Docs}.
\newblock \URLprefix
  \url{https://developer.mozilla.org/en-US/docs/Web/API/OffscreenCanvas}.
  \bibinfo{note}{accessed 2026-07-25}.
\bibitem[{Nejati and Balasubramanian(2016)}]{nejati2016mobile}
\bibinfo{author}{Nejati, J.}, \bibinfo{author}{Balasubramanian, A.},
  \bibinfo{year}{2016}.
\newblock \bibinfo{title}{An in-depth study of mobile browser performance}, in:
  \bibinfo{booktitle}{Proceedings of the 25th International Conference on World
  Wide Web}, \bibinfo{publisher}{ACM}. pp. \bibinfo{pages}{1305--1315}.
\newblock \DOIprefix\doi{10.1145/2872427.2883014}.
\bibitem[{Netravali et~al.(2016)Netravali, Goyal, Mickens and
  Balakrishnan}]{netravali2016polaris}
\bibinfo{author}{Netravali, R.}, \bibinfo{author}{Goyal, A.},
  \bibinfo{author}{Mickens, J.}, \bibinfo{author}{Balakrishnan, H.},
  \bibinfo{year}{2016}.
\newblock \bibinfo{title}{Polaris: Faster page loads using fine-grained
  dependency tracking}, in: \bibinfo{booktitle}{13th USENIX Symposium on
  Networked Systems Design and Implementation}, \bibinfo{publisher}{USENIX
  Association}. pp. \bibinfo{pages}{123--136}.
\bibitem[{Pockstaller et~al.(2023)Pockstaller, Huber and
  Demetz}]{pockstaller2023wasmenergy}
\bibinfo{author}{Pockstaller, D.}, \bibinfo{author}{Huber, S.},
  \bibinfo{author}{Demetz, L.}, \bibinfo{year}{2023}.
\newblock \bibinfo{title}{Comparing the energy consumption of {WebAssembly} and
  {JavaScript} in mobile browsers}, in: \bibinfo{booktitle}{Proceedings of the
  19th International Conference on Web Information Systems and Technologies},
  \bibinfo{publisher}{SciTePress}. pp. \bibinfo{pages}{121--127}.
\newblock \DOIprefix\doi{10.5220/0012205600003584}.
\bibitem[{Procaccianti et~al.(2016)Procaccianti, Fern{\'a}ndez and
  Lago}]{procaccianti2016jss}
\bibinfo{author}{Procaccianti, G.}, \bibinfo{author}{Fern{\'a}ndez, H.},
  \bibinfo{author}{Lago, P.}, \bibinfo{year}{2016}.
\newblock \bibinfo{title}{Empirical evaluation of two best practices for
  energy-efficient software development}.
\newblock \bibinfo{journal}{Journal of Systems and Software}
  \bibinfo{volume}{117}, \bibinfo{pages}{185--198}.
\newblock \DOIprefix\doi{10.1016/j.jss.2016.02.029}.
\bibitem[{Resig(2014)}]{twojs2014particles}
\bibinfo{author}{Resig, J.}, \bibinfo{year}{2014}.
\newblock \bibinfo{title}{{Two.js} particle sandbox}.
\newblock \bibinfo{howpublished}{Chrome Experiments / Google Experiments}.
\newblock \URLprefix
  \url{https://experiments.withgoogle.com/twojs-particle-sandbox}.
  \bibinfo{note}{accessed 2026-07-25}.
\bibitem[{Richards et~al.(2010)Richards, Lebresne, Burg and
  Vitek}]{richards2010analysis}
\bibinfo{author}{Richards, G.}, \bibinfo{author}{Lebresne, S.},
  \bibinfo{author}{Burg, B.}, \bibinfo{author}{Vitek, J.},
  \bibinfo{year}{2010}.
\newblock \bibinfo{title}{An analysis of the dynamic behavior of {JavaScript}
  programs}, in: \bibinfo{booktitle}{Proceedings of the 31st ACM SIGPLAN
  Conference on Programming Language Design and Implementation},
  \bibinfo{publisher}{ACM}. pp. \bibinfo{pages}{1--12}.
\newblock \DOIprefix\doi{10.1145/1806596.1806598}.
\bibitem[{Ruamviboonsuk et~al.(2017)Ruamviboonsuk, Netravali, Uluyol and
  Madhyastha}]{ruamviboonsuk2017vroom}
\bibinfo{author}{Ruamviboonsuk, V.}, \bibinfo{author}{Netravali, R.},
  \bibinfo{author}{Uluyol, M.}, \bibinfo{author}{Madhyastha, H.V.},
  \bibinfo{year}{2017}.
\newblock \bibinfo{title}{Vroom: Accelerating the mobile web with server-aided
  dependency resolution}, in: \bibinfo{booktitle}{Proceedings of the Conference
  of the ACM Special Interest Group on Data Communication},
  \bibinfo{publisher}{ACM}. pp. \bibinfo{pages}{390--403}.
\newblock \DOIprefix\doi{10.1145/3098822.3098851}.
\bibitem[{Selakovic and Pradel(2016)}]{selakovic2016jsperf}
\bibinfo{author}{Selakovic, M.}, \bibinfo{author}{Pradel, M.},
  \bibinfo{year}{2016}.
\newblock \bibinfo{title}{Performance issues and optimizations in {JavaScript}:
  An empirical study}, in: \bibinfo{booktitle}{Proceedings of the 38th
  International Conference on Software Engineering}, \bibinfo{publisher}{ACM}.
  pp. \bibinfo{pages}{61--72}.
\newblock \DOIprefix\doi{10.1145/2884781.2884829}.
\bibitem[{Smits(2020)}]{smits2020performance}
\bibinfo{author}{Smits, A.}, \bibinfo{year}{2020}.
\newblock \bibinfo{title}{Performance, Interactivity, and Collaboration}.
\newblock Ph.D. thesis. Northeastern University.
\newblock \DOIprefix\doi{10.17760/d20401822}. \bibinfo{note}{discusses
  OffscreenCanvas for visualization systems}.
\bibitem[{Surma(2018)}]{surma2018offscreencanvas}
\bibinfo{author}{Surma}, \bibinfo{year}{2018}.
\newblock \bibinfo{title}{{OffscreenCanvas}---speed up your canvas operations
  with a web worker}.
\newblock \bibinfo{howpublished}{web.dev}.
\newblock \URLprefix \url{https://web.dev/articles/offscreen-canvas}.
  \bibinfo{note}{accessed 2026-07-25}.
\bibitem[{{The Chromium Projects}(2014)}]{chromiumJank}
\bibinfo{author}{{The Chromium Projects}}, \bibinfo{year}{2014}.
\newblock \bibinfo{title}{Anatomy of jank}.
\newblock \bibinfo{howpublished}{chromium.org}.
\newblock \URLprefix
  \url{https://www.chromium.org/developers/how-tos/trace-event-profiling-tool/anatomy-of-jank/}.
  \bibinfo{note}{accessed 2026-07-25}.
\bibitem[{{W3C}(2024)}]{longtasks}
\bibinfo{author}{{W3C}}, \bibinfo{year}{2024}.
\newblock \bibinfo{title}{Long tasks {API}}.
\newblock \bibinfo{howpublished}{W3C Working Draft / Editor's Draft}.
\newblock \URLprefix \url{https://w3c.github.io/longtasks/}.
  \bibinfo{note}{accessed 2026-07-25}.
\bibitem[{Wang et~al.(2013)Wang, Balasubramanian, Krishnamurthy and
  Wetherall}]{wang2013wprof}
\bibinfo{author}{Wang, X.S.}, \bibinfo{author}{Balasubramanian, A.},
  \bibinfo{author}{Krishnamurthy, A.}, \bibinfo{author}{Wetherall, D.},
  \bibinfo{year}{2013}.
\newblock \bibinfo{title}{Demystifying page load performance with {WProf}}, in:
  \bibinfo{booktitle}{10th USENIX Symposium on Networked Systems Design and
  Implementation}, \bibinfo{publisher}{USENIX Association}. pp.
  \bibinfo{pages}{473--485}.
\bibitem[{Watt(2018)}]{watt2019mechanising}
\bibinfo{author}{Watt, C.}, \bibinfo{year}{2018}.
\newblock \bibinfo{title}{Mechanising and verifying the {WebAssembly}
  specification}, in: \bibinfo{booktitle}{Proceedings of the 7th ACM SIGPLAN
  International Conference on Certified Programs and Proofs},
  \bibinfo{publisher}{ACM}. pp. \bibinfo{pages}{53--65}.
\newblock \DOIprefix\doi{10.1145/3167082}.
\bibitem[{{web.dev}(2024)}]{inp2024}
\bibinfo{author}{{web.dev}}, \bibinfo{year}{2024}.
\newblock \bibinfo{title}{Interaction to next paint ({INP})}.
\newblock \bibinfo{howpublished}{web.dev}.
\newblock \URLprefix \url{https://web.dev/articles/inp}.
  \bibinfo{note}{accessed 2026-07-25}.
\bibitem[{{WHATWG}(2024)}]{whatwg2024workers}
\bibinfo{author}{{WHATWG}}, \bibinfo{year}{2024}.
\newblock \bibinfo{title}{{HTML Living Standard} --- web workers}.
\newblock \bibinfo{howpublished}{WHATWG}.
\newblock \URLprefix \url{https://html.spec.whatwg.org/multipage/workers.html}.
  \bibinfo{note}{accessed 2026-07-25}.

\end{thebibliography}

\end{document}